# Splay Nematic Phase

Alenka Mertelj,[1] Luka Cmok,[1] Nerea Sebastián,[1] Richard J. Mandle,[2] Rachel R. Parker,[2]
Adrian C. Whitwood,[2] John W. Goodby,[2] and Martin Čopič[1,3]

[1]*Jožef Stefan Institute, SI-1000 Ljubljana, Slovenia*
[2]*Department of Chemistry, University of York, York, YO10 5DD, United Kingdom*
[3]*Faculty of Mathematics and Physics, University of Ljubljana, SI-1000 Ljubljana, Slovenia*

Different liquid crystalline phases with long-range orientational but not positional order, so-called nematic phases, are scarce. New nematic phases are rarely discovered, and such an event inevitably generates much interest. Here, we describe a transition from a uniaxial to a novel nematic phase characterized by a periodic splay modulation of the director. In this new nematic phase, defect structures not present in the uniaxial nematic phase are observed, which indicates that the new phase has lower symmetry than the ordinary nematic phase. The phase transition is weakly first order, with a significant pretransitional behavior, which manifests as strong splay fluctuations. When approaching the phase transition, the splay nematic constant is unusually low and goes towards zero. Analogously to the transition from the uniaxial nematic to the twist-bend nematic phase, this transition is driven by instability towards splay orientational deformation, resulting in a periodically splayed structure. And, similarly, a Landau-de Gennes type of phenomenological theory can be used to describe the phase transition. The modulated splay phase is biaxial and antiferroelectric.

## I. INTRODUCTION

Liquid crystals (LCs) exhibit a rich variety of phases between isotropic liquids and solid states. Until recently, the least ordered nematic phases were an exception, and only a few were observed. Nonchiral molecules are known to form only a simple uniaxial phase, while chiral molecules, in addition to a twisted (chiral) nematic phase, also form three "blue" phases with a lattice of line defects. The very interesting possibility of biaxial nematic thermotropic LC has eluded conclusive experimental discovery so far, in spite of extended efforts by many research groups [1,2].

Recently, it was shown that some bent molecular dimers have a very interesting transition from a normal nematic phase to a conical twisted nematic phase, exhibiting chiral symmetry breaking [3–5]. This transition is driven by an instability towards bent orientational deformation, i.e., pronounced softening of the bend elastic constant [6,7]. As it is impossible to fill the space with a purely bend structure without defects, the spontaneous deformation is both twisted and bent, forming the twist-bend phase.

In the usual uniaxial nematic phase, elongated molecules are oriented, on average, along a common direction, called the director, which is denoted by a unit vector **n** with inversion symmetry $\mathbf{n} \equiv -\mathbf{n}$. Such a nematic phase is not polar, and the order parameter of the phase is a traceless tensor $\mathbf{Q} = S(\mathbf{n} \otimes \mathbf{n} - \frac{1}{3}\underline{\mathbf{I}})$, where $S$ is the scalar order parameter describing how well the molecules are oriented along **n**. The liquid crystalline behavior is strongly influenced by the anisotropic shape of the constituent molecules. While the ground state of the uniaxial nematic is homogeneous, in the chiral nematic LC, the director forms a twisted configuration caused by the chirality of the molecules, whereas in the twist-bend nematic phase, the molecular bend results in the heliconical structure.

We investigate a recently designed polar, rodlike, liquid-crystalline material, which exhibits two distinct nematic mesophases separated by a weakly first-order transition [8,9]. The molecules of the compounds, in which the new nematic phase has been observed so far, have a terminal nitro group on one side and a short terminal chain on the other side. They also possess a large dipole moment and a lateral "bulky group," which gives the molecule a wedge shape. Small-angle x-ray scattering (SAXS) shows a continuous change in the nearest-neighbor correlations rather than a jump at the transition between the nematic mesophases. In this paper, we show that this transition is analogous to the twist-bend transition but with the splay as the unstable deformation, probably due to the wedge shape of the constituent molecules. As it is impossible to fill the space with homogeneous splay, the deformed structure exhibits a periodically modulated splay. We study the transition between the nematic phases using polarization microscopy, dynamic light scattering, and wide-angle x-ray scattering (WAXS) and SAXS. In Sec. II, we present the experimental results. In Sec. III, we discuss the results and propose a Landau-de Gennes type of phenomenological theory to explain the observed behavior.

## II. EXPERIMENTAL RESULTS

The liquid-crystalline material used in our study (RM734) exhibits an isotropic-to-nematic phase transition at $T_{IN} = 187.9\,°C$, a nematic-to-second-nematic transition at $T_{NNs} = 132.7\,°C$, and a melting point at $T_m = 139.8\,°C$. In the following, we denote the lower nematic phase as the $N_S$ phase. Although the transition to the second nematic phase appears below the melting point, we could perform measurements of the $N$-$N_S$ transition with a cooling rate down to 0.03 K/min. The synthesis of RM734 is described in Ref. [9]; however, we subjected the material to further purification to attempt to remove trace ionic impurities as follows. RM734 (1 g) was dissolved into anhydrous dichloromethane (20 ml) and washed with saturated aqueous ethylenediaminetetraacetic acid disodium salt (10 ml). The aqueous layer was discarded, the organic dried over $MgSO_4$ and filtered over basic alumina (50 g), eluting with anhydrous DCM (300 ml). The solvents were removed *in vacuo* to give a white solid; this was recrystallized twice from distilled ethyl acetate affording a microcrystalline white solid.

The B3LYP/6-31G(d) minimized geometry of RM734 was determined using the Gaussian G09e01 suite of programs [10] and is shown in Fig. 1. It exhibits a large dipole moment of 11.3748 D [calculated by the B3LYP/6-31G(d) level of DFT], which is oriented almost along the molecular long axis.

The material was filled into either planar LC cells (with rubbed surfaces to induce orientation of **n** parallel to the cell surface) or homeotropic LC cells (with the surfaces treated to induce perpendicular orientation of **n**). The cells were filled in either a higher-temperature nematic or an isotropic phase.

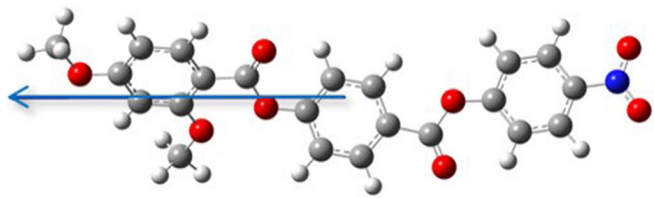

FIG. 1. The B3LYP/6-31G(d) minimized geometry of RM734 used in this study. The blue arrow shows the orientation of the dipole moment, which is oriented almost along the molecular long axis ($x$: 11.3031 D, $y$: 0.7132 D, $z$: −1.0572 D).

### A. Polarization microscopy

First, we studied the transition between the nematic phases using polarization microscopy. In thick planar cells (Instec, 20 $\mu$m), the director **n** was well aligned in the plane of the cell along the rubbing direction [Fig. 2(a)]. To better observe the features of the transition, the sample was rotated so that the angle between the analyzer and **n** was about 20°. During slow cooling (0.1 K/min), strong fluctuations were observed just above the phase transition destabilizing the homogeneous orientation of **n** [Fig. 2(b)]. After the homogeneous structure was restored [Fig. 2(c)], a white line traveled across the sample (Fig. 2(d), Movie 1 [11]). The $N_S$ phase was well oriented and appeared the same as the $N$ phase. When observed on heating, the $N$-$N_S$ transition looked similar to that observed on cooling, except for the white line [Fig. 2(d)], which is only observed upon cooling.

Fast cooling (the heater was switched off) resulted in a more inhomogeneous transition and the creation of defects (Fig. 3, Movie 2 [11]). In the $N_S$ phase, thick lines and loops decorated with pointlike defects were observed. The loops slowly shrank and annihilated. Some of the lines remained, probably because their relaxation was hindered by pinning to the spacers and other imperfections of the sample. During fast heating, no additional defects appeared. The thick lines from the $N_S$ phase became even thicker in the $N$ phase and split into two thin lines, probably disclinations (Movies 2 and 3 [11]). This suggests that the thick lines in the $N$ phase may be $\pm 1$ defect lines or walls, which are not stable in this phase, so they split into two $\pm 1/2$ lines.

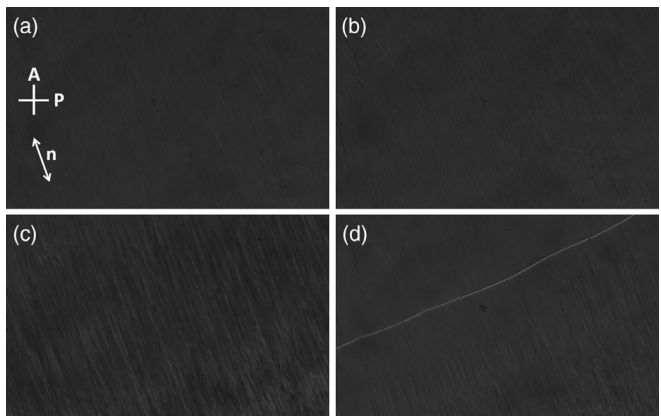

FIG. 2. Polarization microscopy images showing the $N$-$N_S$ phase transition at a cooling rate of 0.1 K/min in a planar cell (thickness 20 $\mu$m). (a) Homogeneously oriented sample in the $N$ phase, (b) strong splay fluctuations just above the $N$-$N_S$ phase transition, (c) destabilization of the homogeneous orientation during the transition, and (d) restoration of the homogeneous order in the $N_S$ phase. The orientation of **n**, polarizer P, and analyzer A are shown in (a). For better observations, the sample was rotated so that the angle between **n** and the analyzer was around 20°. The width of the images is 545 $\mu$m.

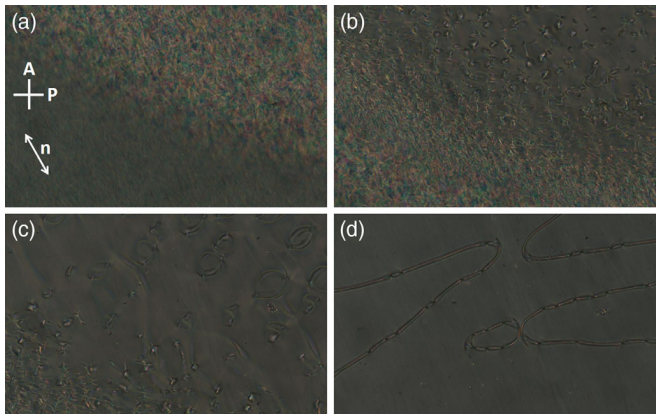

FIG. 3. Polarization microscopy images showing the $N$-$N_S$ phase transition during fast cooling in a planar cell (thickness 20 $\mu$m). (a) Destabilization of the nematic order, (b,c) evolution of defects, and (d) defects in the $N_S$ phase. The orientation of **n**, the polarizer, and the analyzer are shown in (a). For better observations, the sample was rotated so that the angle between **n** and the analyzer was around 20°. The width of the images is 545 $\mu$m.

In a thick homeotropic cell (Instec, 18 $\mu$m), we were not able to achieve homogeneous ordering of **n** in either of the nematic phases. A schlieren texture was observed in both phases (Fig. 4); however, the $N_S$ phase exhibited a significantly larger number of defect lines. In comparison to the $N$ phase, after initial relaxation, two additional defect lines originating from the centers of $\pm 1$ defects and an additional defect line originating from the centers of $\pm 1/2$ defects were observed [Fig. 4(b)]. This indicates that the symmetry of the $N_S$ phase is different (lower) than that of the ordinary $N$ phase.

Polarization microscopy was also used for measurement of anisotropy of the index of refraction $\Delta n$. A sample in a planar cell with thickness $d$ was rotated between crossed polarizers so that the angle between the polarizer and the director was 45 degrees. The intensity of monochromatic light transmitted through the sample was measured to determine the phase difference $\phi = 2\pi \Delta n d/\lambda$ between the ordinary and extraordinary light. An interference filter for $\lambda = 632.8$ nm was used to filter the light from the halogen lamp in the microscope. In the $N$ phase, $\Delta n$ is proportional to the scalar order parameter $S$. Figure 5 shows the temperature dependence of $\Delta n$. First, it increases with decreasing temperature, as expected in the usual nematic phase, where the average orientational order of the molecules increases with decreasing temperature. In the region about 2 K above the $N$-$N_S$ phase transition, where strong fluctuations appear, $\Delta n$ begins to decrease. At the transition, the measurements show a slight increase of $\Delta n$, but this is probably an artefact due to light scattering from fluctuations. The measured value of $\Delta n$ in the $N_S$ phase just below the transition is about 0.01 smaller than that in the $N$ phase close to the transition, before the fluctuations cause its decrease. In the $N_S$ phase, it decreases with decreasing temperature.

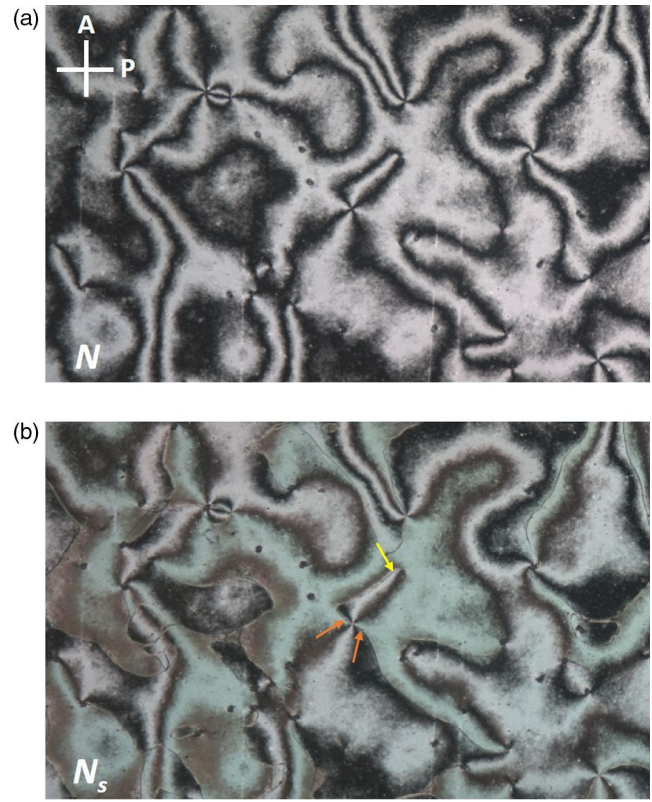

FIG. 4. Polarization microscopy images in (a) $N$ (135 °C) and (b) $N_S$ (125 °C) phases in a homeotropic cell (thickness 18 $\mu$m). The appearance of two additional defect lines around $\pm 1$ defects and one around $\pm 1/2$ defects in the $N_S$ phase indicates that the symmetry of this phase is lower than that of the $N$ phase. The polarizer and analyzer are oriented as shown in (a). The width of the images is 545 $\mu$m. The yellow and orange arrows point at additional lines around one $\pm \frac{1}{2}$ and one $\pm 1$ defect, respectively.

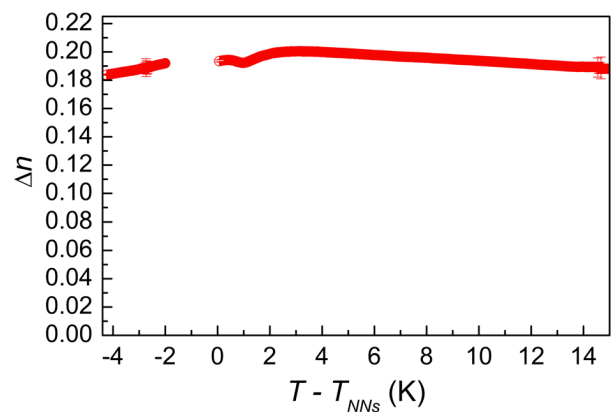

FIG. 5. Temperature dependence of anisotropy of index of refraction. In the temperature region where the orientation is not homogeneous, we were not able to measure $\Delta n$.

### B. Dynamic light scattering

In order to understand what drives the transition to the second nematic phase, we measure the director orientational fluctuations in the vicinity of the $N$-$N_S$ phase transition by dynamic light scattering (DLS). From these measurements, we obtain the temperature dependence of the elastic constants and some viscosity coefficients. We measure the scattered intensity and relaxation rates of eigenmodes of orientational fluctuations. By choosing the correct scattering geometry (Appendix A), pure modes can be measured with intensities $I_i \propto (\Delta\varepsilon_{\mathrm{opt}})^2/K_i q^2$ and relaxation rates $1/\tau_i = K_i q^2/\eta_i$, where $i = 1, 2, 3$ denote splay, twist, and bend, and $q$ is the scattering vector [12]. While twist viscosity equals rotational viscosity $\gamma_1$, $\eta_2 = \gamma_1$, the values of bend and splay viscosities are affected by backflow, $\eta_1 = \gamma_1 - \alpha_3^2/\eta_b$ and $\eta_3 = \gamma_1 - \alpha_2^2/\eta_c$, where $\alpha_i$ are Leslie viscosity coefficients and $\eta_{b,c}$ Miesowicz viscosities [12]. In the case of the splay, the reduction of orientational viscosity due to the backflow is usually small (of the order of a few %), so $\eta_1 \approx \gamma_1$. The temperature dependence of the anisotropy of the dielectric tensor at optical frequencies $\Delta\varepsilon_{\mathrm{opt}}$ is obtained from measurements of $\Delta n$ (Fig. 5). With this method, the temperature dependence of elastic constants is obtained but not their absolute values. We determine the absolute values of $K_1$ and $K_3$ at 150 °C from measurement of Fredericks transition (see below), and $K_2$ from the ratio between $K_1/\eta_1$ and $K_2/\eta_2$ at 145 °C, i.e., assuming that, in the case of splay, the backflow is negligible.

Ratios $K_i/\eta_i$, obtained from the measured relaxation rates of pure fluctuation modes, are shown in Fig. 6. When approaching the $N$-$N_S$ phase transition, a strong slowing down of splay and twist modes is observed, while the bend mode becomes only slightly slower. In the $N_S$ phase, two modes are observed in the splay and the bend geometry. The value of the relaxation rate of the slower splay mode is similar to the splay mode in the $N$ phase before slowing down; the second mode is of about 1 order of magnitude faster. The relaxation rate of the first bend mode is similar to the bend mode in the $N$ phase, while the second is of 1 order of magnitude slower. On the other hand, only one mode is detected for the twist geometry, where the relaxation rate jumps to a value slightly smaller than in the $N$ phase before slowing down.

Figure 7(a) shows the temperature dependence of the elastic constants close to the $N$-$N_S$ phase transition. The most prominent feature is that the splay elastic constant $K_1$ is unusually low and goes toward zero when approaching the phase transition. In the observed temperature range, the twist elastic constant $K_2$ and the bend elastic constant $K_3$ slowly increase with decreasing temperature, as expected for usual $S^2$ dependence. Additionally, the rotational viscosity $\gamma_1$ increases steeply when approaching the transition [Fig. 7(b)]. It should be noted that the temperature dependence of the splay viscosity is the same as that of the rotational viscosity, and thus, the assumption that the first is not significantly affected by the backflow is justified. The bend viscosity slightly increases when approaching the phase transition.

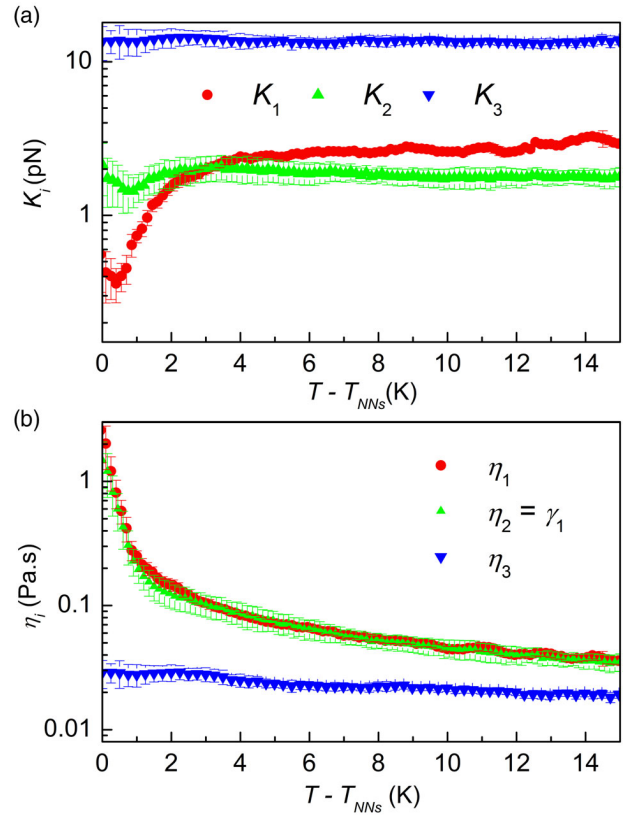

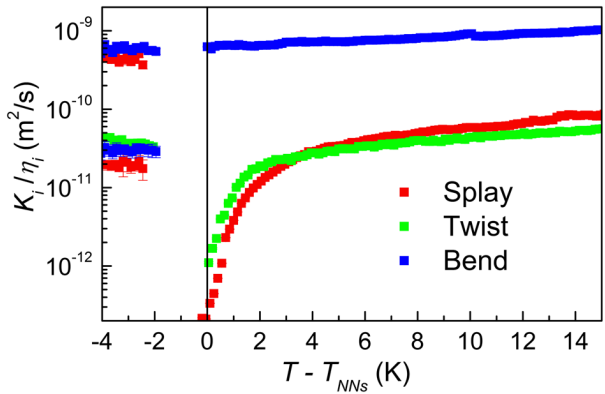

FIG. 6. Temperature dependence of ratios $K_i/\eta_i$, which are proportional to the relaxation rates of orientational fluctuations. The cooling rate is 0.03 K/min. In the temperature region where the orientation is not homogeneous, we were not able to reliably determine $K_i/\eta_i$.

FIG. 7. Temperature dependence of (a) elastic constants $K_i$ and (b) viscosities $\eta_i$ in the $N$ phase close to the $N$-$N_S$ phase transition. The most interesting is the behavior of the splay elastic constant, which goes toward zero when approaching the phase transition. The cooling rate is 0.03 K/min.

### C. Dielectric measurements

The reference absolute values for $K_1$ and $K_3$ used to calculate the temperature dependence of the elastic constants from the scattered intensity data were obtained from dielectric and optical measurements of the Fredericks transition. At a given temperature, the nematic dielectric spectra shows two pronounced relaxations, one at around 1 kHz and the other one a little below 1 MHz, which shifts down for large probe voltages. The first relaxation is related to a strong electrode polarization effect introduced by the polyimide alignment layer, while the latter one corresponds to the characteristic cell relaxation provoked by the finite resistivity of ITO electrodes. Optical observation during the spectra measurements at large voltages shows that at low frequencies the applied field is screened by the space charged regions created in the vicinity of the electrodes and polyimide layers due to the diffusion of ion impurities, and no reorientation of the director is achieved. However, clear homeotropic orientation is achieved at frequencies of 10 kHz, at lower frequencies than ITO relaxation, and thus, Fredericks transition was studied at this frequency. Dielectric and optical Fredericks curves are given in Fig. 8 together with their corresponding fits for a temperature 20 degrees above the phase transition. Both values for the splay elastic constant $K_1$, obtained from the dielectric and optical measurements, are in reasonably good agreement; 2 pN and 2.4 pN, respectively. The determination of the value of $K_3$ is less reliable. The fit in the case of dielectric measurements is significantly better than the optical one, so we took its results for reference values of the bend elastic constant, which is 11.6 pN. We should note here that the obtained ratio $K_3/K_1$ is about 6. This ratio is in good agreement with the estimated value for $K_3/K_2$ of 5 obtained from DLS (Appendix A).

Measurements were also performed at temperatures closer to the $N$-$N_S$ transition; however, the large drop of the observed threshold voltage made an unambiguous determination of the constants very difficult. Although these values are tentative, it should be stressed that the error on the determination of the splay and bend elastic constants will only result in a shift up or down of the data plotted in Fig. 7 but not of their relative behavior and temperature dependence.

The observed large values of dielectric permittivity deserve further notice and can possibly be attributed to strong dipolar correlations already in the $N$ phase. Unfortunately, the sample shows rather large conductivity effects; thus, detailed dielectric spectroscopy analyses are left out of the scope of this paper. Nevertheless, all these data suggest an exceptional dielectric behavior of the material, which deserves a thorough study by itself, provided that the conductivity problems are overcome.

### D. SAXS and WAXS

We subjected RM734 to a SAXS/WAXS study using CuK$_\alpha$ radiation ($\lambda = 1.54184$ Å) at variable temperatures. Both the $N$ and $N_S$ phases exhibit only diffuse scattering at small and wide angles, refuting the possibility of either nematic phase being a lamellar (smectic) phase.

Both nematic phases exhibit multiple scattering events parallel to the aligning magnetic field ($B$); therefore, radial integration (to give scattered intensity versus Q) was performed on the wedge segments identified in Fig. 9(a). The scattering profile parallel to $B$ was deconvoluted into three peaks by fitting with Lorentzians, allowing the d-spacing of the (001), (002), and (003) peaks to be determined [Figs. 9(b) and 9(c)]. The spacing of the (002) and (003) peaks is effectively temperature invariant, whereas the (001) peak is largely temperature invariant in the nematic phase but decreases continuously in the $N_S$ phase [Fig. 9(d)]. The (001) peak has a maximum value of 21 Å in the $N$ phase and a minimum of 19.1 Å in the $N_S$ phase; these values compare with a molecular length of 20.8 Å and indicate that neither phase experiences significant molecular interdigitation (antiparallel overlap). Only a single scattering peak (identified as 100) is observed perpendicular to $B$; this corresponds to the average lateral molecular separation, and this exhibits a continual decrease in both $N$ and $N_S$ phases; see Fig. 13 in Appendix C. Other than the small change in d-spacing of the (001) peak, we do not observe any significant differences (peak intensity, d-spacing, FWHM, etc.) in the SAXS/WAXS patterns of the $N$ and $N_S$ phases.

Although demonstrated by DFT(B3LYP/6-31G(d)) optimized geometry, we obtained a single crystal of RM734 suitable for structure determination by x-ray diffraction (XRD) by the vapor diffusion technique using ethyl acetate and cyclohexane. Because of the lateral methoxy group, the molecular breadth of RM734 is greater at one end than at the other, and thus RM734 can be described as being slightly wedge shaped (Fig. 10). Full details are given in Appendix D.

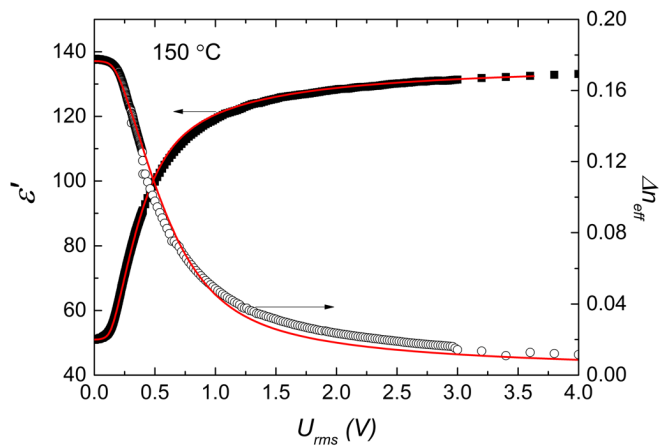

FIG. 8. Dielectric (full squares) and optical (empty circles) Fredericks transition at 150 °C and 10 kHz. The lines are the fits.

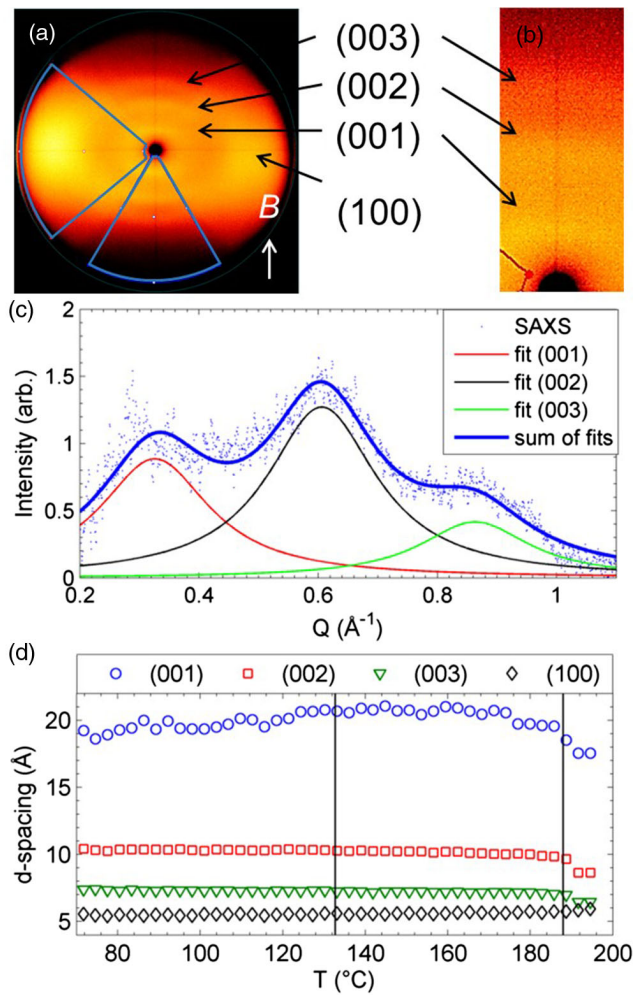

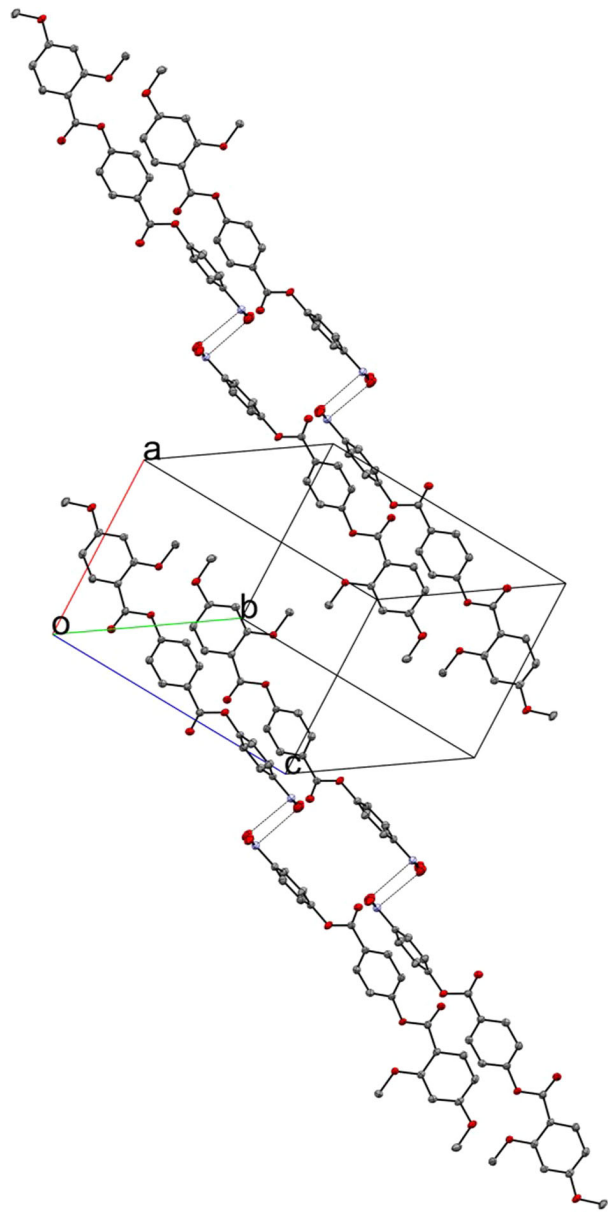

FIG. 9. (a) 2D SAXS pattern of RM734 in the $N_S$ phase at 130 °C; the 2D scattering data are radially integrated in the regions defined by the two wedges in (a) to give d-spacings of the (001), (002), and (003) peaks [bottom wedge, see Fig. 9(c)] and the (100) peak (left wedge). (b) Expansion of the small-angle region of (a) showing three weak scattering events. (c) Fitting of raw SAXS data at 80 °C to extract d-spacings of the (001), (002), and (003) peaks. (d) Plot of the d-spacings of small- [(001), (002), (003)] and wide-angle [(100)] peaks as a function of temperature. Solid lines correspond to phase transitions.

FIG. 10. Structure of RM734 in the solid state displayed as a thermal ellipsoid model (50% probability), obtained via x-ray diffraction as described in the text and Appendix D. The unit cell (space group $P\bar{1}$) is indicated. Dashed lines correspond to nitro-nitro close contacts, as described in the text.

We observe noninterdigitated antiparallel pair formation in the solid state (Fig. 10) via a nitro-nitro close contact with N(O)O-N(O)$_2$ separations, of which there are two per pair, of 3.085(2) Å and 3.104(2) Å, i.e., slightly greater than the van der Waals radii. As this interaction is specific for nitro- containing materials, we propose that the pair formation via this motif is essential for the $N_S$ phase, and we note that other polar groups that pair form (e.g., CN) do not show this phase [9].

Within each individual molecule, we observe no coplanarity of aromatic rings, with each phenyl-phenyl ring dihedral angle of about approximately 48°, giving each molecule (and thus antiparallel pair) a twisted ribbon structure. We note that this structure is not apparent in the DFT(B3LYP/6-31G(d)) optimized geometry, this being the only significant difference between the optimized structure and that obtained by XRD.

## IV. DISCUSSION AND THEORY

The results show that the low-temperature nematic mesophase has lower symmetry than the high-temperature one. The temperature behavior of the splay elastic constant is very similar to the case of the transition from the nematic to the twist-bend nematic phase, where the bend elastic constant goes toward zero. As a consequence, in the twist-bend phase, periodic spontaneous bend deformation accompanied by a small twist appears [6,7]. Analogously, we may expect that the splay constant going towards zero will cause the appearance of spontaneous splay. Similarly, the structure should be periodic because a homogeneous splay deformation cannot fill the space without defects. The simplest periodic splayed configuration is shown in Fig. 11. However, the $N_S$ phase observed by polarization microscopy is homogeneous, so if there is periodic structure, its period is smaller than the wavelength of light. Periodic structure with small periods would cause a decrease of anisotropy of index of refraction, which we observed in the $N_S$ phase (Fig. 5). The length corresponding to the (001) peak [Fig. 9(d)] decreases in the $N_S$ phase as would be expected for the tilted packing depicted in Fig. 11, supporting results obtained from the measurement of $\Delta n$ and suggesting a temperature dependence in the splay angle.

In a manner similar to that used to explain the transition between the nematic and the twist-bend phase [6,13], we can describe the $N$-$N_S$ transition by a Landau–de Gennes type of free energy. The most appropriate way to describe a nematic LC is in terms of tensor order parameter **Q**. In the usual $N$ phase, it is uniaxial and can be written in the form $\mathbf{Q} = S(\mathbf{n} \otimes \mathbf{n} - \frac{1}{3}\mathbf{I})$. To model the $N$-$N_S$ transition, only the part of free energy due to the elastic deformation is needed, i.e., expansion to the necessary order in $\nabla \mathbf{Q}$. To the lowest order $(\nabla \mathbf{Q})^2$, the splay and bend elastic constants are equal. This degeneracy is broken by the invariants of the form $\mathbf{Q}(\nabla \mathbf{Q})^2$, three of which are relevant. In the nematic phase, splay deformation gives rise to flexoelectric polarization, so we assume that the $N$-$N_S$ transition is driven by flexoelectric coupling of polarization **P** and $\nabla \mathbf{Q}$. This mechanism was introduced by Shamid et al. [13] to model the transition to the twist-bend phase. Thus, the free energy density with a minimum number of terms necessary for the $N$-$N_S$ transition is

$$f_{el} = \frac{1}{2}C_1 Q_{ij,k}Q_{ij,k} - C_5 Q_{ij}Q_{ik,k}Q_{jl,l} - \gamma_0 Q_{ij,j}P_i$$
$$+ \frac{1}{2}tP_i P_i + \frac{1}{2}bP_{i,j}P_{i,j}. \quad (1)$$

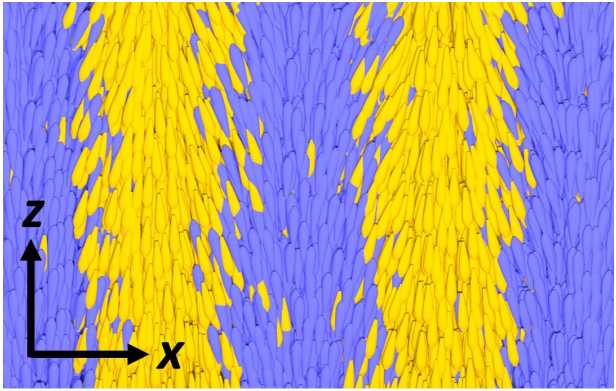

FIG. 11. Schematic of the proposed splay phase, where different colors denote orientation of the molecules either along **n** or −**n**.

Here, $\gamma_0$ is a bare flexoelectric coefficient. We take $t$ to decrease with decreasing temperature, driving the transition. Note that the $C_5$ term breaks the degeneracy of $K_1$ and $K_3$: $K_1 = C_1 S^2 - \frac{4}{3}C_5 S^3$ and $K_3 = C_1 S^2 + \frac{2}{3}C_5 S^3$. We assume that $C_5 > 0$, so $K_3 > K_1$. The term $P_{i,j}P_{i,j}$ is necessary to stabilize the splayed phase. Other terms are either not relevant to the transition to the splayed state or they give the same expression, when the parametrized deformation is inserted into Eq. (1).

For the splayed state, we seek a solution of the form $\mathbf{n} = (\sin(\vartheta), 0, \cos(\vartheta))$ and $\mathbf{P} = P\cos(kx)\mathbf{n}$, where $\vartheta = \vartheta_0 \sin(kx)$ (see Appendix E). We assume that **P** and **n** are parallel because the dipole moments of the molecules are along the molecular long axis. Assuming $S$ is constant, we insert the solution into Eq. (1), then expand the free energy to fourth order in $\vartheta_0$ and $P$, and average it over the period $2\pi/k$,

$$f_s = \frac{1}{2}K_1 k^2 \vartheta_0^2 - \gamma k \vartheta_0 P k + \frac{1}{2}tP^2 + \frac{1}{2}bP^2 k^2$$
$$+ \frac{1}{2}bP^2 \vartheta_0^2 k^2 + \frac{1}{8}tP^2 \vartheta_0^2 + \frac{1}{4}C_5 S^3 k^2 \vartheta_0^4. \quad (2)$$

Here, $\gamma = \gamma_0 S$. The fifth and sixth terms are of second order in coupling between **P** and $\nabla \mathbf{Q}$, so we omit them. (It can be shown by numerical calculation that these higher-order terms affect the solution very little.) The minimization of Eq. (2) then yields

$$P = \frac{\gamma k \vartheta_0}{t + bk^2}, \quad (3)$$

which, when inserted into Eq. (2), results in

$$f_s = \frac{1}{2}\left(K_1 - \frac{\gamma^2}{t + bk^2}\right)k^2 \vartheta_0^2 + \frac{1}{4}C_5 S^3 k^2 \vartheta_0^4. \quad (4)$$

Above the phase transition, $t$ is larger than the critical value $t_c = \gamma^2/K_1$, and the minimization gives $k = 0$. The effective splay elastic constant can be written as $K_{1\text{eff}} = K_1 - \gamma^2/t$, which is positive and so $f_s > 0$. Thus, the uniform nematic state ($f_s = 0$) is favorable. When $t$ reaches the critical value $t_c$, $f_s < 0$, and the splayed configuration with $k \neq 0$ becomes more stable in comparison to the uniform state.

It is interesting that by changing the sign of $C_5$ in the free energy density [Eq. (1)], the structure becomes unstable towards bend deformation, which leads to the twist-bend or splay-bend phase. The coupling term renormalizes both $K_1$ and $K_3$, and since the $C_5$ term enters the expressions for $K_1$ and $K_3$ with different signs, the sign of $C_5$ determines which elastic deformation becomes unstable.

Taking $\Delta t = t_c - t$, the parameters just below the phase transition to the lowest order in $\Delta t$ can be expressed as

$$\vartheta_0(\Delta t) = \sqrt{\frac{2K_1 \Delta t}{3C_5 S^3 \gamma^2}}$$

$$k(\Delta t) = \sqrt{\frac{\Delta t}{3b}}$$

$$P(\Delta t) = \frac{K_1^2 \Delta t}{3\gamma^2} \sqrt{\frac{2}{bC_5 S^3}}. \quad (5)$$

As the $N_S$ phase appears optically homogeneous, the deformation period $2\pi/k$ must be in the nanometer range, similarly to the twist-bend phase. Thus, at optical scale, the $N_S$ phase must be biaxial. The change of $\Delta n$ at the $N$-$N_S$ transition (Fig. 5) depends on the direction of spontaneous splay and is connected to the tilt of the molecules. In the $xz$ plane (Fig. 11), the change is about approximately $-\Delta n\vartheta_0^2$, and in the $yz$ plane, it is about approximately $-\Delta n\vartheta_0^2/2$; thus, the difference between the in-plane indices of refraction $n_y$ and $n_x$, which are equal in the uniaxial nematic phase, is about $-\Delta n\vartheta_0^2/2$. The measured change of $\Delta n$ at the phase transition is about approximately 0.01, so the amplitude of the tilt $\vartheta_0$ can be estimated to be about 10°. Therefore, $n_y - n_x$ is also of the order of 0.01. Because of the modulated splay, as a result of the flexoelectric polarization, the $N_S$ phase is also antiferroelectric.

Nonresonant SAXS measurements did not show any signature of periodic structure (below a limit of Q approximately 0.05 Å$^{-1}$, $d$ approximately 125 Å). The periodicity of a purely splay modulated nematic phase as depicted in Fig. 11 lacks electron density modulation (EDM) along the modulation axis and, as such, should *not* be observed in nonresonant SAXS experiments.

The $N$-$N_S$ transition is similar to the (splay) Fredericks transition, which is a field-induced, second-order, structural transition in a layer with thickness $d$, at which one splay fluctuation (with $k \sim \pi/d$) exhibits critical slowing down and leads the transition from a uniform to a deformed structure. At the $N$-$N_S$ transition, all fluctuations slow down, and, in principle, this transition could be of second order, with $k$ increasing from 0 at the transition to a finite, temperature-dependent value [Eq. (5)]. In a real system, the sample size $L$ would limit the smallest possible $k$ at the phase transition to about approximately $2\pi/L$. The theory could easily be modified to describe a first-order phase transition by adding higher-order terms in $\mathbf{P}$ to the free energy in Eq. (1). A full tensor analysis of possible modulated nematic structures that arises from coupling between molecular steric polarization and nematic order has been studied by Longa and Pająk [14]. However, in their model, they did not include the third-order elastic terms [$C_5$ in Eq. (1)], which breaks the degeneracy of $K_1$ and $K_3$. Among the predicted periodic structures, there is also transverse periodic nematic phase, which, in the case that $S$ is constant, reduces to the structure we propose for the splay phase.

In a LC material made of pear- or wedge-shaped molecules, a splay modulation is also accompanied by a modulation of the proportion of the molecules, which on average point in the $\mathbf{n}$ and $-\mathbf{n}$ directions. This mechanism, on one hand, results in a splay flexoelectricity in the $N$ phase and in an antiferroelectric order in the $N_S$ phase. On the other hand, it softens the splay elastic constant, as predicted, e.g., by Gregorio et al. [15] in systems made of conical constituents. The correlation between shape, polar order, and splay was theoretically demonstrated in several cases. A lattice model for splay ferroelectricity in a system of uniaxial pear-shaped molecules predicts a phase diagram showing isotropic, nematic, and polar phases, for which polar order is accompanied by splay [16]. Phases with spontaneous splay deformation with defects were predicted in polar nematic liquid crystal [17,18]. However, until now, experimental realization of splayed phases remained elusive. On the other hand, softening of the splay elastic constant has been observed experimentally. There are reports of a decrease of the splay elastic constant in materials, when bent-shape molecules are added to the nematic phase of rod molecules [19], although adding bent-shape molecules usually results in a decrease of the bend elastic constant [20,21]. It has been shown that a lateral group is required for materials to exhibit the $N$-$N_S$ transition [9]. This group gives the molecule a wedge shape and increases its biaxiality. The XRD study shows that, in the crystal, the molecules with parallel orientation of the dipole moments have lateral groups, oriented in the same plane, and their positions are shifted along the long axis (Fig. 10). Such an average configuration of the neighboring molecules is what is expected to occur during the splay deformation, giving rise to flexoelectric polarization and biaxial structure.

So far, the $N$-$N_S$ transition has been observed in materials whose molecules have a nitro group, the reason for which is currently unclear. XRD on crystal structure shows that a nitro-nitro close contact leads to formation of nonintercalated, tail-to-tail, antiparallel pairs. In a liquid, such pairing is a dynamic process, so we can speculate that, in the $N_S$ phase, and during a splay fluctuation in the $N$ phase, the probability for pair formation would be the largest in the regions where the polarization changes sign (Fig. 11). In principle, this would cause a slight modulation of viscosity. Because of the rigidity of the bond and, consequently, the biaxiality of the pairs, the pairing may be a mechanism that promotes biaxiality, which might be one of the key prerequisites for the splay phase.

## V. CONCLUSIONS

We have shown that the second, low-temperature nematic phase observed in recently designed polar, rodlike LC materials has lower symmetry than the ordinary uniaxial nematic phase. It exhibits defect structures that are usually not observed in the uniaxial nematic phase. The $N$-$N_S$ phase transition is weakly first order, with a significant pretransitional behavior, which manifests as strong splay fluctuations. In the $N$ phase, the splay nematic constant is unusually low and goes towards zero when approaching the phase transition. Analogously to the transition from the uniaxial nematic to the twist-bend nematic phase, the $N$-$N_S$ phase transition is driven by an instability towards splay orientational deformation, resulting in a periodically splayed structure. Similarly, a Landau-de Gennes type of phenomenological theory can be used to describe the phase transition. Optically, the $N_S$ phase looks homogeneous and very similar to the $N$ phase, which indicates that the splay period is smaller than the wavelength of light. So again, as in the case of the twist-bend phase, additional studies—e.g., resonant x-ray [22] or freeze fraction transmission electron microscopy [4,5]—could help to determine the period. We proposed the simplest structure for the splay phase, which is supported by the experiments. The proposed structure is biaxial and antiferroelectric, which agrees with the observed symmetry. Formation and classification of topological defects would certainly be an interesting future study.

## ACKNOWLEDGMENTS

A. M. and M. Č. acknowledge financial support from the Slovenian Research Agency (A. M. and M. Č. from Research Core Funding No. P1-0192, and A. M. from Project No. J7-8267). N. S. thanks the European Union's Horizon 2020 Research and Innovation Programme for its support through the Marie Curie Individual Fellowship No. 701558 (MagNem). L. C. acknowledge support from MIZŠ&ERDF funds OPTIGRAD Project. R. J. M. and J. W. G. acknowledge funding from EPSRC (UK) for the Bruker SAXS/SWAXS equipment via EP/K039660/1 and for ongoing work via EP/M020584/1.

## APPENDIX A: DYNAMIC LIGHT SCATTERING

In the DLS experiments, we used a standard setup, with a frequency-doubled diode-pumped ND:YAG laser (532 nm, 80 mW), an ALV APD based "pseudo" cross-correlation detector, and an ALV-6010/160 correlator to obtain the autocorrelation function of the scattered light intensity. The direction and the polarization of the incoming and detected light were chosen so that pure modes were observed (Fig. 12). A single-mode optical fiber with a GRIN lens was used to collect the scattered light within one coherence area [23]. We fitted the intensity autocorrelation function $g_2$ with $g_2 = 1 + 2(1-j_d)j_d g_1 + j_d^2 g_1^2$, where $j_d$ is the ratio between the intensity of the light that is scattered inelastically and the total scattered intensity, and $g_1$ was either a single (in the $N$ phase), $g_1 = \mathrm{Exp}(-t/\tau)$, or a double exponential function, $g_1 = a_1\mathrm{Exp}(-t/\tau_1) + (1-a_1)\mathrm{Exp}(-t/\tau_2)$ (in some cases, in the $N_S$ phase). The relaxation rate $1/\tau$ was attributed to the chosen eigenmode of orientational fluctuations with the wavevector $\mathbf{q}$ equal to the scattering vector $\mathbf{q_s}$. The scattered intensity of a given mode was determined as a product $j_d I_{\mathrm{tot}}$, where $I_{\mathrm{tot}}$ was the total detected intensity.

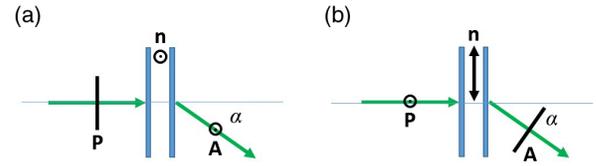

FIG. 12. DLS scattering geometries. (a) Splay ($\alpha = 35°$) and twist ($\alpha = 3°$), and (b) bend ($\alpha = 35°$).

By measuring the angular dependence in the geometry shown in Fig. 12(b), the dependence of the twist-bend relaxation rate of fluctuations on the ratio between components of scattering vectors perpendicular ($\perp$) and parallel ($\parallel$) to $\mathbf{n}$ can be obtained:

$$\frac{1}{\tau} = \frac{K_2 q_\perp^2 + K_3 q_\parallel^2}{\gamma_1 - \frac{\alpha_2^2 q_\parallel^2}{\eta_a q_\perp^2 + \eta_c q_\parallel^2}}. \tag{A1}$$

Thus, it is possible to evaluate the ratio $K_3/K_2$. The method is not very sensitive, but it gives a good estimate, in particular, when $K_3/K_2 \gg 1$. At 150 °C, the measurements gave $K_3/K_2 \sim 5$.

## APPENDIX B: DIELECTRIC MEASUREMENTS

Determination of the splay and bend elastic constants was carried out by means of the capacitance method using 8 $\mu$m antiparallel planar aligned cells. A planar-to-homeotropic Fredericks transition was induced by applying an ac signal with an Agilent Precision LRC meter E4980A. The capacitance of the sample was measured as a function of the applied voltage, which was varied from 0.001 Vrms to 10 Vrms, with a delay time of 20 s between the application of the ac signal and the acquisition of the capacitance value. Simultaneously, the transmitted intensity of the sample between crossed polarizers (obtained by using an interference filter with $\lambda = 632.8$ nm to filter the halogen lamp of the microscope) was recorded with a CMOS camera (IDS Imaging UI-3370CP). Values of the splay and bend elastic constants were extracted from the fitting of the entire voltage dependence of the capacitance to the theory [24].

## APPENDIX C: SAXS/WAXS

SAXS/WAXS was performed on a Bruker D8 diffractometer (Cu-K$_\alpha$ radiation, $\lambda = 1.54184$ Å) equipped with a bored graphite rod furnace, affording temperature control to ±0.1 °C. The sample was confined in a capillary (0.9 mm ID) and aligned with a pair of 1T magnets perpendicular to the incident beam. Raw SAXS/WAXS data are available upon request from the University of York data catalogue.

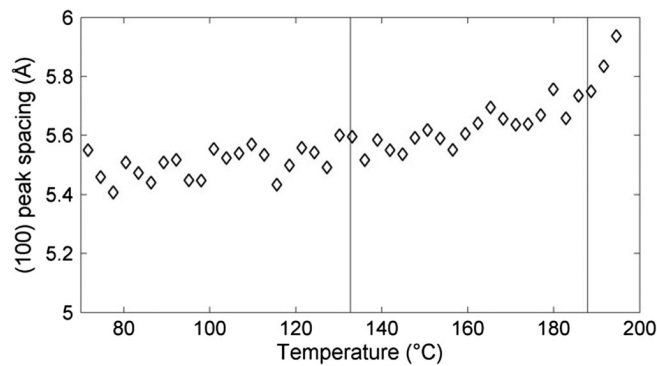

FIG. 13. Plot of the d-spacing of the (100) peak (wide angle) as a function of temperature, where solid lines correspond to the locations of the I-N and N-$N_S$ phase transitions, respectively.

The (100) peak captures information regarding the average lateral molecular separation, which we find to decrease almost linearly across both the N and $N_S$ phases. Solid lines indicate the temperatures of phase transitions, which were determined by DSC; while we can identify the I-N transition with ease by SAXS/WAXS, there is effectively no change in the SAXS/WAXS pattern when going from the N to the $N_S$ phase. The d-spacing of the (100) peak was determined by fitting raw scattering data at each temperature studied with a Lorentzian function (Fig. 13).

## APPENDIX D: SINGLE-CRYSTAL X-RAY DIFFRACTION (XRD)

Single crystals of $C_{22}H_{17}NO_8$ (RM734) were grown from ethyl acetate by slow vapor diffusion of cyclohexane (about 1 week). Diffractometer control, data collection, initial unit-cell determination, frame integration, and unit-cell refinement were carried out with "Crysalis" [25]. Face-indexed absorption corrections were applied using spherical harmonics, implemented in the SCALE3 ABSPACK scaling algorithm [26].

Single-crystal diffraction data were collected on an Oxford Diffraction SuperNova diffractometer with Cu-Kα radiation ($\lambda = 1.54184$ Å) using an EOS CCD camera. The crystal was kept at 110.05(10) K during data collection. Using Olex2 [27], the structure was solved with the ShelXT [28] structure solution program using intrinsic phasing and refined with the ShelXL [29] refinement package using least squares minimization.

*Crystal Data* for $C_{22}H_{17}NO_8$ ($M = 423.36$ g/mol): triclinic, space group P-1 (no. 2), $a = 11.2452(6)$ Å, $b = 11.4399(5)$ Å, $c = 16.0975(7)$ Å, $\alpha = 70.817(4)°$, $\beta = 80.085(4)°$, $\gamma = 79.626(4)°$, $V = 1909.60(17)$ Å$^3$, $Z = 4$, $T = 110.05(10)$ K, $\mu(CuK\alpha) = 0.962$ mm$^{-1}$, Dcalc = 1.473 g/cm$^3$, 12 608 reflections measured ($8.052 \leq 2\Theta \leq 134.148°$), and 6832 unique ($R_{int} = 0.0170$, $R_{sigma} = 0.0246$), which were used in all calculations. The final $R_1$ was 0.0353 [$I > 2\sigma(I)$], and $wR_2$ was 0.0967 (all data). The unit cell is shown in Fig. 14. The structure is deposited with the CCDC, deposition number 1851381.

Our motivation to grow single crystals of RM734 was, aside from providing an unambiguous demonstration of the molecular structure of this material, to study the possibility of antiparallel pair formation, which, given the propensity of nitro-containing materials to form dynamic antiparallel pairs in bulk liquid-crystal phases [30], we considered quite likely. SAXS/WAXS experiments on alkyl/alkoxy cyanobiphenyls show that the (001) peak has a far larger d-spacing than the molecular length due to interdigitation (i.e., partial overlap) of the antiparallel pairs, which can also display a temperature dependency [31]. However, for RM734, the d-spacing of the (001) peak is roughly equal to the molecular length in the N and $N_S$ phases. In the solid state, we observe that RM734 forms noninterdigitated antiparallel pairs, which are the consequence of a close contact between nitro groups on adjacent molecules (see Fig. 10). The N(O)O—N(O)$_2$ separations, of which there are two per pair, are 3.085(2) Å and 3.104(2) Å, which is in the range of the van der Waals contact distance and thus implies a possible noncovalent nitro-nitro interaction. However, we note that this separation is larger than the Bondi–van der Waals radii [32], so we cautiously describe this as a close contact throughout. If we label the rings of each molecule as A, B, and C (or A′, B′, and C′ for molecules in opposite orientations) (Fig. 15), we can clearly identify ABC-C′B′A′ pairs as being parallel to other A′B′C′-CBA chains as shown in Fig. 15. Note how the orientation of the para-methoxy group alternates between parallel ABC-ABC pairs but remains constant between the antiparallel ABC-A′B′C′.

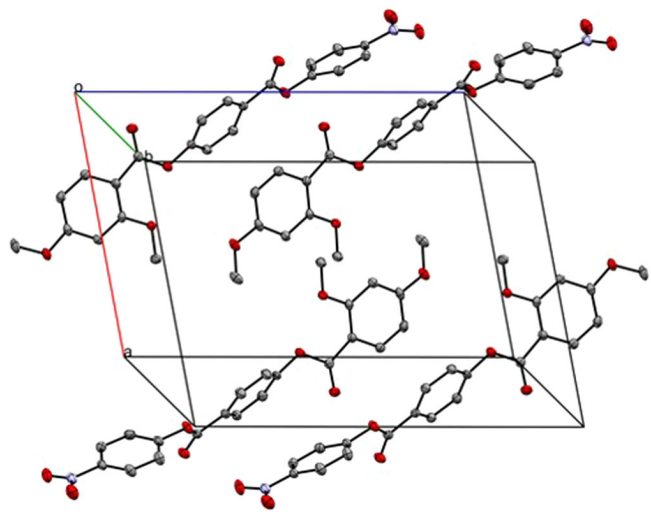

FIG. 14. The unit cell of RM734 ($P\bar{1}$ space group) displayed with a thermal ellipsoid model (50% probability) as obtained by x-ray diffraction on single crystals grown by vapor diffusion of cyclohexane into ethyl acetate.

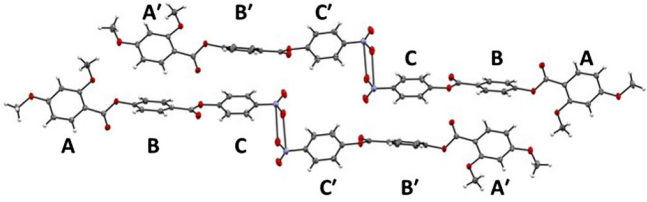

FIG. 15. Structure of RM734 in the solid state. The dashed lines indicate the close contacts between $_{(nitro)}$O—N$_{(nitro)}$ in adjacent molecules (two per pair). Rings are labeled A, B, C (A′, B′, C′ on adjacent pair) as per discussion in the text.

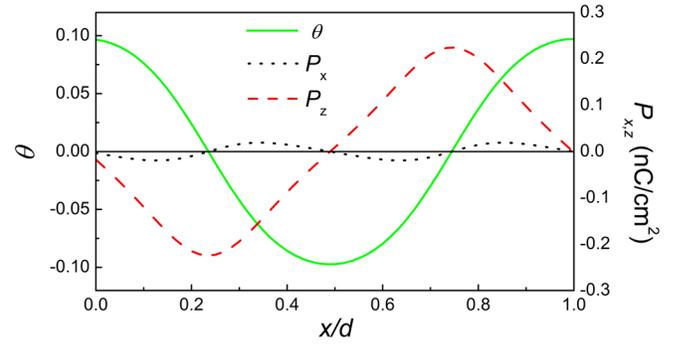

FIG. 16. Numerical solution of the global minimization of the free energy in 1D for $t = 0.98 t_c$. The sample size is varied to obtain the solution, which corresponds to the minimum of the energy. For $C_1 S^2 = 8$ pN, $C_5 S^3 = 4$ pN, $\gamma = 2.8 \times 10^{-6}$ V, and $b = 4 \times 10^{-5}$ Vm/(As).

Previously in Ref. [9], we reported that a nitro-terminated material is essential for the incidence of the $N_S$ phase, as replacement with other polar groups (CN, SF$_5$) presumably prevents formation of this phase since these groups do not support the formation of an offset noninterdigitated pair. This interaction (with comparable $_{(nitro)}$O—N$_{(nitro)}$ separation distances) has been reported previously in several other systems [33–36].

## APPENDIX E: STABILITY ANALYSIS OF NEMATIC PHASE

In order to show that the Ansatz for the splay phase describes the correct fluctuation, which causes the transition, we performed stability analysis of the nematic phase described by the free energy density given in Eq. (1) against small periodic fluctuations $\vartheta = \vartheta_0 e^{-ik \cdot x} + \text{c.c.}$ and $P_{x,z} = P_{0x,0z} e^{-ik \cdot x} + \text{c.c.}$ in $\mathbf{n}(x) = [\sin(\vartheta(x)), 0, \cos(\vartheta(x))]$ and $\mathbf{P} = (P_x(x), 0, P_z(x))$. The free energy density [Eq. (1)] was expanded to second order in $\vartheta(x)$, $P_x(x)$, and $P_z(x)$, averaged over $x$, and minimized with respect to $\vartheta_0$ and $P_{0x,0z}$. In this order, $P_x(x)$ is not coupled to $\vartheta(x)$, and for positive $t$ and $b$, minimization gives $P_{0x} = 0$. $P_{0z}$ and $\vartheta_0$ are coupled, and minimization yields two branches of fluctuation modes with eigenvectors

$$(\vartheta, P_z)_{1,2} = (\pm \vartheta_{1,2} \sin(kx), P_{z1,z2} \cos(kx)) \quad \text{(E1)}$$

and eigenvalues

$$\lambda_{1,2} = \frac{1}{4}\left( bk^2 + K_1 k^2 + t \mp \sqrt{b^2 k^4 - 2bK_1 k^4 + K_1^2 k^4 + 2bk^2 t - 2K_1 k^2 t + t^2 + 4k^2 \gamma^2} \right). \quad \text{(E2)}$$

As long as the eigenvalues are positive, the nematic phase is stable towards the fluctuations. The first eigenvalue $\lambda_1$ becomes negative for $t < \gamma^2/K_1 - bq^2$, so this is the fluctuation branch that destabilizes the nematic phase and leads the transition.

We also performed a global numerical minimization of the free energy [Eq. (1)] as a function of the sample size, which yields similar results; that is, for $t < t_c$, a periodic modulation of the type used for the Ansatz minimizes the free energy (Fig. 16).

[1] G. R. Luckhurst and T. J. Sluckin, *Final Remarks*, in *Biaxial Nematic Liquid Crystals* (Wiley-Blackwell, Chichester, West Sussex, 2015), pp. 369–374.

[2] M. Lehmann, *Low Molar Mass Thermotropic Systems*, in *Biaxial Nematic Liquid Crystals* (Wiley-Blackwell, Chichester, West Sussex, 2015), pp. 333–367.

[3] M. Cestari, S. Diez-Berart, D. A. Dunmur, A. Ferrarini, M. R. de la Fuente, D. J. B. Jackson, D. O. Lopez, G. R. Luckhurst, M. A. Perez-Jubindo, R. M. Richardson, J. Salud, B. A. Timimi, and H. Zimmermann, *Phase Behavior and Properties of the Liquid-Crystal Dimer 1″,7″-bis(4-cyanobiphenyl-4′-yl) Heptane: A Twist-Bend Nematic Liquid Crystal*, Phys. Rev. E **84**, 031704 (2011).

[4] D. Chen, J. H. Porada, J. B. Hooper, A. Klittnick, Y. Shen, M. R. Tuchband, E. Korblova, D. Bedrov, D. M. Walba, M. A. Glaser, J. E. Maclennan, and N. A. Clark, *Chiral Heliconical Ground State of Nanoscale Pitch in a Nematic Liquid Crystal of Achiral Molecular Dimers*, Proc. Natl. Acad. Sci. U.S.A. **110**, 15931 (2013).

[5] V. Borshch, Y.-K. Kim, J. Xiang, M. Gao, A. Jákli, V. P. Panov, J. K. Vij, C. T. Imrie, M. G. Tamba, G. H. Mehl, and

O. D. Lavrentovich, *Nematic Twist-Bend Phase with Nanoscale Modulation of Molecular Orientation*, Nat. Commun. **4**, 2635 (2013).

[6] I. Dozov, *On the Spontaneous Symmetry Breaking in the Mesophases of Achiral Banana-Shaped Molecules*, Europhys. Lett. **56**, 247 (2001).

[7] K. Adlem, M. Čopič, G. R. Luckhurst, A. Mertelj, O. Parri, R. M. Richardson, B. D. Snow, B. A. Timimi, R. P. Tuffin, and D. Wilkes, *Chemically Induced Twist-Bend Nematic Liquid Crystals, Liquid Crystal Dimers, and Negative Elastic Constants*, Phys. Rev. E **88**, 022503 (2013).

[8] R. J. Mandle, S. J. Cowling, and J. W. Goodby, *A Nematic to Nematic Transformation Exhibited by a Rod-like Liquid Crystal*, Phys. Chem. Chem. Phys. **19**, 11429 (2017).

[9] R. J. Mandle, S. J. Cowling, and J. W. Goodby, *Rational Design of Rod-like Liquid Crystals Exhibiting Two Nematic Phases*, Chem. Eur. J. **23**, 14554 (2017).

[10] M. J. Frisch, G. W. Trucks, H. B. Schlegel, G. E. Scuseria, M. A. Robb, J. R. Cheeseman, G. Scalmani, V. Barone, G. A. Petersson, H. Nakatsuji, X. Li, M. Caricato, A. Marenich, J. Bloino, B. G. Janesko, R. Gomperts, B. Mennucci, H. P. Hratchian, J. V. Ortiz, A. F. Izmaylov et al., *Gaussian 09, Revision A.02* (Gaussian, Inc., Wallingford, CT, 2016).

[11] See Supplemental Material at http://link.aps.org/supplemental/10.1103/PhysRevX.8.041025 for Movies of the $N$-$N_S$ phase transition and relaxation of defects in a planar cell (thickness 20 $\mu$m). The width of the images is 545 $\mu$m. The number in the upper-right corner denotes the rate at which we speed up the movie. Movie 1: Polarization microscopy showing the $N$-$N_S$ phase transition at a cooling rate of 0.1 K/min. Movie 2: Polarization microscopy images showing the $N$-$N_S$ phase transition during fast cooling. Movie 3: Decay of a thick defect line/wall in two thin lines in the $N$ phase.

[12] P. G. de Gennes and J. Prost, *The Physics of Liquid Crystals*, 2nd ed. (Clarendon Press, Oxford, 1995).

[13] S. M. Shamid, S. Dhakal, and J. V. Selinger, *Statistical Mechanics of Bend Flexoelectricity and the Twist-Bend Phase in Bent-Core Liquid Crystals*, Phys. Rev. E **87**, 052503 (2013).

[14] L. Longa and G. Pająk, *Modulated Nematic Structures Induced by Chirality and Steric Polarization*, Phys. Rev. E **93**, 040701 (2016).

[15] P. D. Gregorio, E. Frezza, C. Greco, and A. Ferrarini, *Density Functional Theory of Nematic Elasticity: Softening from the Polar Order*, Soft Matter **12**, 5188 (2016).

[16] S. Dhakal and J. V. Selinger, *Statistical Mechanics of Splay Flexoelectricity in Nematic Liquid Crystals*, Phys. Rev. E **81**, 031704 (2010).

[17] H. Pleiner and H. R. Brand, *Spontaneous Splay Phases in Polar Nematic Liquid Crystals*, Europhys. Lett. **9**, 243 (1989).

[18] S. M. Shamid, D. W. Allender, and J. V. Selinger, *Predicting a Polar Analog of Chiral Blue Phases in Liquid Crystals*, Phys. Rev. Lett. **113**, 237801 (2014).

[19] J.-H. Lee, T.-H. Yoon, and E.-J. Choi, *Unusual Temperature Dependence of the Splay Elastic Constant of a Rodlike Nematic Liquid Crystal Doped with a Highly Kinked Bent-Core Molecule*, Phys. Rev. E **88**, 062511 (2013).

[20] P. Sathyanarayana, M. Mathew, Q. Li, V. S. S. Sastry, B. Kundu, K. V. Le, H. Takezoe, and S. Dhara, *Splay Bend Elasticity of a Bent-Core Nematic Liquid Crystal*, Phys. Rev. E **81**, 010702 (2010).

[21] S. Parthasarathi, D. S. S. Rao, K. F. Csorba, and S. K. Prasad, *Viscoelastic Behavior of a Binary System of Strongly Polar Bent-Core and Rodlike Nematic Liquid Crystals*, J. Phys. Chem. B **118**, 14526 (2014).

[22] C. Zhu, M. R. Tuchband, A. Young, M. Shuai, A. Scarbrough, D. M. Walba, J. E. Maclennan, C. Wang, A. Hexemer, and N. A. Clark, *Resonant Carbon K-Edge Soft X-Ray Scattering from Lattice-Free Heliconical Molecular Ordering: Soft Dilative Elasticity of the Twist-Bend Liquid Crystal Phase*, Phys. Rev. Lett. **116**, 147803 (2016).

[23] T. Gisler, H. Rüger, S. U. Egelhaaf, J. Tschumi, P. Schurtenberger, and J. Rička, *Mode-Selective Dynamic Light Scattering: Theory versus Experimental Realization*, Appl. Opt. **34**, 3546 (1995).

[24] S. W. Morris, P. Palffy-Muhoray, and D. A. Balzarini, *Measurements of the Bend and Splay Elastic Constants of Octyl-Cyanobiphenyl*, Mol. Cryst. Liq. Cryst. **139**, 263 (1986).

[25] CrysAlisPro, Oxford Diffraction Ltd. Version 1.171.34.41 (n.d.).

[26] Empirical absorption correction using spherical harmonics, implemented in the SCALE3 ABSPACK scaling algorithm within CrysAlisPro software, Oxford Diffraction Ltd. Version 1.171.34.40 (n.d.).

[27] O. V. Dolomanov, L. J. Bourhis, R. J. Gildea, J. a. K. Howard, and H. Puschmann, *OLEX2: A Complete Structure Solution, Refinement and Analysis Program*, J. Appl. Crystallogr. **42**, 339 (2009).

[28] G. M. Sheldrick, *SHELXT—Integrated Space-Group and Crystal-Structure Determination*, Acta Crystallogr. Sect. A **71**, 3 (2015).

[29] G. M. Sheldrick, *Crystal Structure Refinement with SHELXL*, Acta Crystallogr. Sect. C **71**, 3 (2015).

[30] R. J. Mandle, S. J. Cowling, I. Sage, M. E. Colclough, and J. W. Goodby, *Relationship between Molecular Association and Re-entrant Phenomena in Polar Calamitic Liquid Crystals*, J. Phys. Chem. B **119**, 3273 (2015).

[31] R. J. Mandle, E. J. Davis, C.-C. A. Voll, D. J. Lewis, S. J. Cowling, and J. W. Goodby, *Self-Organisation through Size-Exclusion in Soft Materials*, J. Mater. Chem. C **3**, 2380 (2015).

[32] A. Bondi, *van der Waals Volumes and Radii*, J. Phys. Chem. **68**, 441 (1964).

[33] R. Soman, D. Raghav, S. Sujatha, K. Rathinasamy, and C. Arunkumar, *Axial Ligand Modified High Valent Tin(IV) Porphyrins: Synthesis, Structure, Photophysical Studies and Photodynamic Antimicrobial Activities on Candida albicans*, RSC Adv. **5**, 61103 (2015).

[34] O. Bolton and A. J. Matzger, *Improved Stability and Smart-Material Functionality Realized in an Energetic Cocrystal*, Angew. Chem. **123**, 9122 (2011).

[35] I. Caracelli, S. H. Maganhi, P. J. S. Moran, B. R. S. de Paula, F. N. Delling, and E. R. T. Tiekink, *Crystal structure of (3E)-3-(2,4-dinitrophenoxymethyl)-4-phenylbut-3-en-2-one*, Acta Crystallogr. Sect. E **70**, o1051 (2014).

[36] M. Daszkiewicz, *Importance of O–N Interaction between Nitro Groups in Crystals*, CrystEngComm **15**, 10427 (2013).